\documentclass[
]{ceurart}

\usepackage{listings}
\usepackage{float}
\begin{document}

\copyrightyear{2025}
\copyrightclause{Copyright for this paper by its authors.
  Use permitted under Creative Commons License Attribution 4.0
  International (CC BY 4.0).}

\conference{ITADATA-WS 2025: The 4$^{\text{th}}$ Italian Conference on Big Data and Data Science -- Workshops, September 9--11, 2025, Turin, Italy}

\title{Exascale In-situ visualization for Astronomy \& Cosmology}

\author[1,2]{Nicola Tuccari}[%
orcid=0009-0004-7802-2602,
email=nicola.tuccari@inaf.it,
]
\cormark[1]
\fnmark[1]
\address[2]{Universit\`{a} di Catania, Dipartimento di Matematica e Informatica, Catania, Italy}

\author[1]{Eva Sciacca}[%
orcid=0000-0002-5574-2787,
email=eva.sciacca@inaf.it,
]
\fnmark[1]
\address[1]{INAF Astrophysical Observatory of Catania, Catania, Italy}

\author[3]{Yolanda Becerra}[%
orcid=0000-0003-2357-7796,
email=yolanda.becerra@bsc.es
]
\fnmark[1]
\address[3]{Barcelona Supercomputing Center, Barcelona, Spain}
\author[3]{Enric Sosa Cintero}[%
email=enric.sosacintero@bsc.es
]
\fnmark[1]
\author[2]{Emiliano Tramontana}[%
orcid=0000-0002-7169-659X,
email=emiliano.tramontana@unict.it
]
\fnmark[1]
\fntext[1]{These authors contributed equally.}

\begin{abstract}
Modern simulations and observations in Astronomy \& Cosmology (A\&C)  produce massively large data volumes, posing significant challenges for storage, access and data analysis. A long-standing bottleneck in high-performance computing, especially now in the exascale era, has been the requirement to write these large datasets to disks, which limits the performance.

A promising solution to this challenge is in-situ processing, where analysis and visualization are performed concurrently with the simulation itself, bypassing the storage of the simulation data. 

In this work, we present new results from an approach for in-situ processing based on Hecuba, a framework that provides a highly distributed database for streaming A\&C simulation data directly into the visualization pipeline to make possible on-line visualization. By integrating Hecuba with the high-performance cosmological simulator ChaNGa, we enable real-time, in-situ visualization of N-body simulation results using tools such as ParaView and VisIVO.
\end{abstract}


\maketitle

\section{Introduction}

Astrophysical observations and simulation codes executed on high-performance supercomputers generate massive data volumes, often reaching petabyte scales. Managing such data poses considerable challenges that must be addressed to enable scientific discoveries\cite{hey2009the}. Emerging pre-exascale infrastructures offer new possibilities for scaling applications in Astronomy and Cosmology (A\&C), in particular to support high-performance visualization, which is essential for understanding simulation outcomes.

Traditionally, scientific visualization has relied on post-processing, where simulation data is first stored permanently on disk and only later retrieved for analysis and visualization. However, this type of workflow can be inefficient given the massive amount of data generated. An alternative processing paradigm is in-situ visualization, which enables the analysis and rendering of data in real time, concurrently to the execution of the simulation, thus allowing users to generate the visualization as the data is generated.

Processing such massive datasets is only feasible with the use of HPC resources. A major bottleneck in achieving high performance and scalability comes from the reliance on parallel file systems. Additionally, file-based workflows often introduce execution patterns that require strict synchronization and complex code, limiting the capability to adapt to new requirements or hardware changes.

A practical and emerging alternative for scientific computing are Key-Value (KV) databases, as they are particularly well suited for handling time-series and spatial datasets. Moreover, they allow to analyze partial results and react, for instance, by discarding a cosmological simulation as soon as a certain event occurs.

By leveraging Hecuba, we enable a specific form of in-situ visualization known as in-transit visualization \cite{moreland2011examples}, where analysis and visualization are executed on dedicated I/O nodes. These I/O nodes receive the full simulation results, but they avoid to store them, but rather they generate new information from analysis or provide run-time visualization. 

In previous work\cite{HPV25}, we presented the first steps of the integration of the Hecuba platform within the ChaNGa high-performance cosmological simulator and the in-situ visualization of its N-body results with the ParaView and VisIVO tools. 
In this paper, we present further developments, focusing on the ParaView plugin and reducing VisIVO dependence on file-based data storing, to enable in-situ visualization.

\section{Related Works}
In-situ visualization has become increasingly important in the field of high-performance computing, as simulations now produce massive volumes of data that are impractical to store and analyze with just post-processing. Several surveys in the literature provide comprehensive overviews of existing techniques and tools for in-situ visualization, highlighting their central role in scientific workflows that demand real-time analysis while reducing I/O overheads \cite{insitu}.

Three main approaches to in-situ visualization can be identified: tighty coupled, loosely coupled and hybrid. In the tightly coupled model, visualization routines are directly embedded within the simulation code, which allows efficient memory access and minimal latency but requires synchronous execution and additional memory resources. The loosely coupled model, by contrast, separates computation and visualization into independent processes that may run on different resources and communicate through shared memory or network interfaces. Finally, hybrid approaches attempt to balance the advantages of both by performing data reduction during the simulation and forwarding the processed results to external resources for further analysis.

Several frameworks have been developed to support these approaches. One example is VisIt Libsim \cite{libsim}, which implements the tightly coupled model by enabling simulation codes to function as a compute engine directly linked to VisIt. Another widely adopted framework is ParaView Catalyst \cite{catalyst}, which provides flexible integration mechanisms within simulation codes and supports both tightly and loosely coupled configurations.

These solutions have demonstrated the feasibility and advantages of in-situ workflows, but they also highlight persistent challenges, such as the complexity of code instrumentation and the risk of resource contention between simulation and visualization tasks.

\section{Data, Tools and Methodology}

\subsection{ChaNGa Cosmological Data}

ChaNGa\footnote{\url{https://github.com/N-BodyShop/changa}} (Charm N-body GrAvity solver) is a parallel code designed for collisionless N-body simulations in astrophysics. It is capable of performing both cosmological simulations with periodic boundary conditions in comoving coordinates and simulating isolated stellar systems. It also includes hydrodynamics modeling using the Smooth Particle Hydrodynamics (SPH) method. Gravity calculations are computed using the Barnes-Hut tree algorithm, which includes hexadecapole expansion of nodes and Ewald summation for handling periodic forces. For time integration, ChaNGa uses a leapfrog integrator that allows each particle to evolve with its own individual timestep. ChaNGa stores data using Tipsy, a binary file format used primarily in astrophysical simulations to store snapshot data of N-body systems, it can store three types of particles: gas, dark matter and star.

\subsection{Hecuba}

Hecuba\footnote{\url{https://github.com/bsc-dd/hecuba}} is a set of tools and interfaces that aim to facilitate the management and utilization of persistent data for Big Data applications.
\begin{figure}[h]
    \centering
    \includegraphics[width=0.8\linewidth]{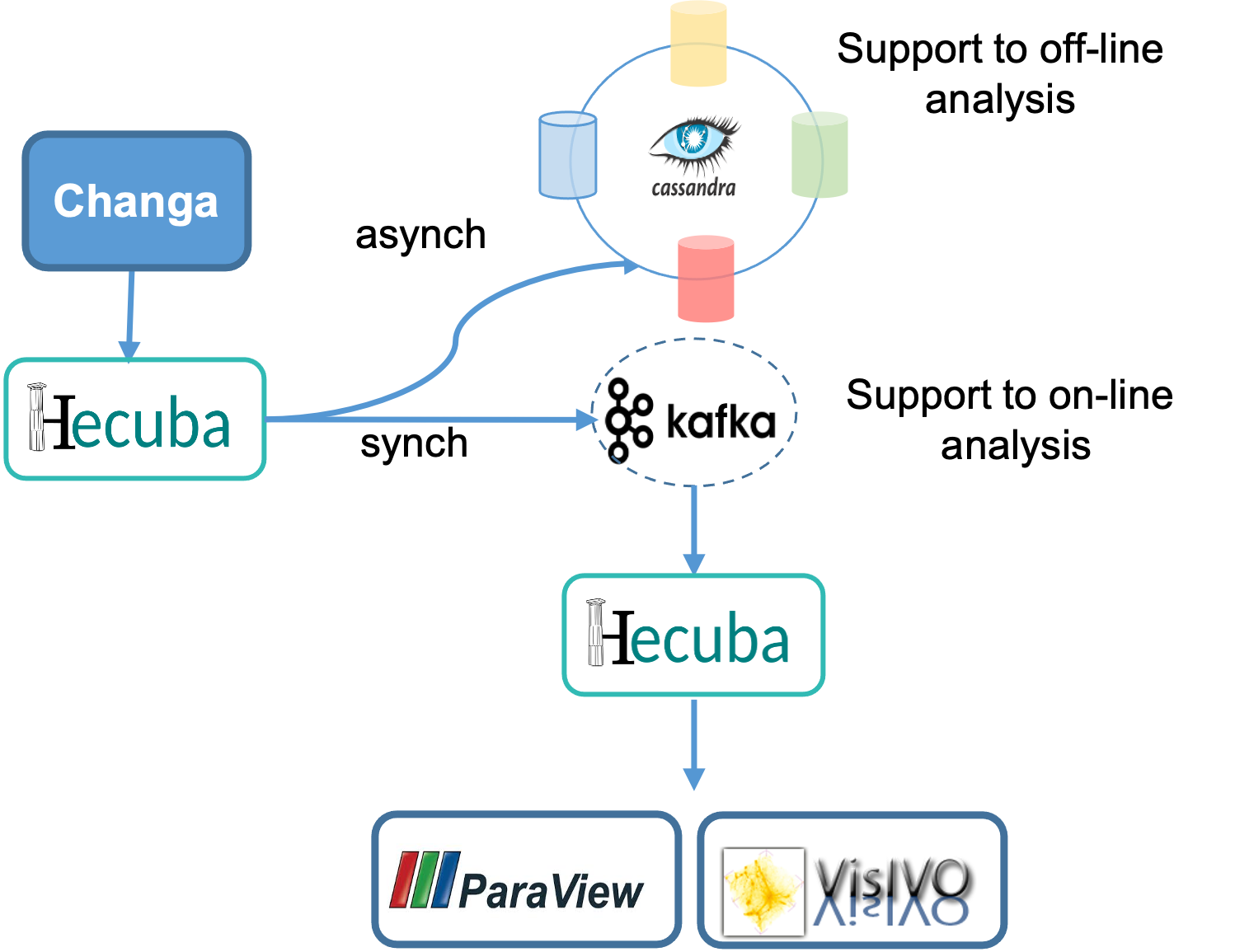}
    \caption{Hecuba architecture implementing an Object Mapper for Apache Cassandra support both off-line and on-line ChaNGa data analysis.}
    \label{fig:enter-label}
\end{figure}

Hecuba implements an Object Mapper for Apache Cassandra \cite{CASS10} (a recognized NoSQL database) that enables programmers to interact with it and its data as regular in-memory objects. Currently, Hecuba implements an interface for Python and C++ programming languages.
In order to provide a mechanism to synchronize run-time visualizations, we are extending Hecuba to implement a lambda architecture integrating Apache Kafka\footnote{Apache Kafka, \url{https://kafka.apache.org/}}. 
Lambda architecture is a data-processing structure defined to speed up online analysis for Big Data applications.
With this approach, Hecuba allows at the same time to persist the generated data in a key-value datastore and also to produce a stream of data for online analysis.
Notice that programmers can use the same interface to access the data, whether it is in memory, in storage, or it is coming through a stream: they only need to modify the class definition of the object to indicate which kind of object it is.

\subsection{ParaView}

ParaView\footnote{\url{https://www.paraview.org/}}\cite{PVIEW05} is a tool designed to support the visualization and analysis of large scientific data sets.
To this end, ParaView supports, for example,  parallel data processing and rendering to enable interactive visualization or remote parallel computing.
The decoupled architecture of ParaView allows it to run the setup as a client-server model with two separate processes: a server which runs on a potentially powerful remote machine and a ParaView client which runs on a desktop machine.
Moreover, ParaView is an extensible framework that uses a complete plugin mechanism to allow the addition of new functionalities.
In this work, we have implemented a custom ParaView plugin that uses the Hecuba interface to implement the online acquisition of the data to visualize. Also, to perform further comparisons for evaluation, we have extended this plugin to provide the option to read data from a Tipsy file.

\subsection{VisIVO}
\begin{figure}[h]
    \centering
    \includegraphics[width=0.8\linewidth]{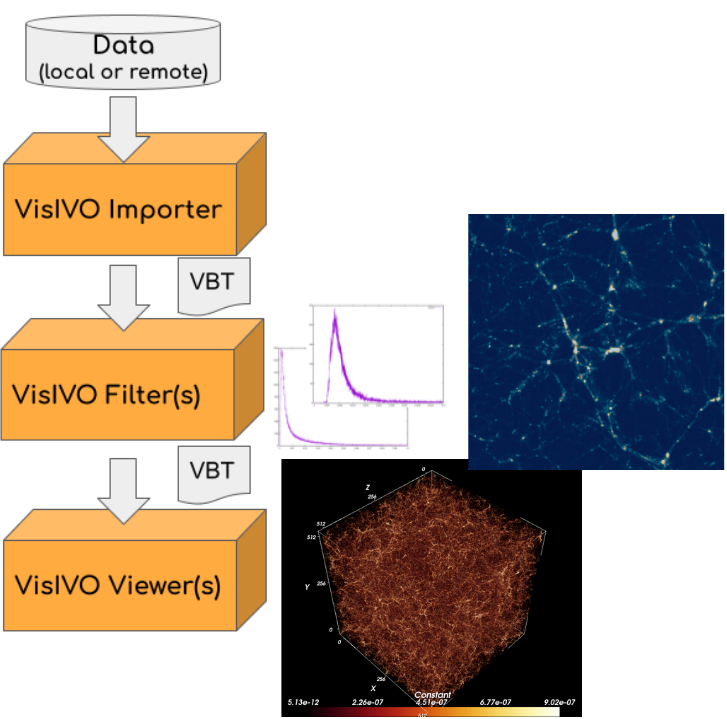}
    \caption{Typical visualization pipeline of VisIVO Server consisting on the application of the three main modules.}
    \label{fig:visivo-workflow}
\end{figure}

VisIVO is a software designed for high-performance, multi-dimensional visualization and analysis of large-scale astrophysical datasets\cite{sciacca2015integrated}. VisIVO Server\footnote{\url{https://github.com/VisIVOLab/VisIVOServer}} stands out as a modular platform designed for creating customized views of 3D renderings from astrophysical datasets. It consists of three core components: VisIVO Importer, VisIVO Filter, and VisIVO Viewer. All of these components make use of the VisIVO Binary Table (VBT) data structure, a highly efficient data representation internally used by VisIVO Server. We are working on evolving VisIVO to take advantage of the capabilities provided by modern high-performance computing environments. A typical visualization pipeline performed with VisIVO Server modules is shown in figure \ref{fig:visivo-workflow}.

We presented the integration of Hecuba within VisIVO Server in our previous work\cite{ADASS2024}. This integration allowed VisIVO Server to interact with a highly distributed database or use its streaming capabilities to perform in-situ visualization. 

To complete the work to make VisIVO compatible with in-situ visualization, we needed to update VisIVO's implementation in order to avoid the reliance on the storage. 

To reach this requirement, as a first step, we extended \texttt{VSTable}, a VisIVO Server internal data structure, to support in-memory data handling. The VSTable data structure is responsible for performing all the required functions to modify or create a VBT. To support  in-memory data handling, we developed a new subclass, \texttt{VSTableMem}, in which all columns are stored in memory using \texttt{std::vector}. Furthermore, all functions responsible for data access and manipulation, such as \texttt{PutColumn} and \texttt{GetColumn}, were redefined to operate on the in-memory data structures. This modification allows VisIVO to avoid relying on VBT files and, at the same time, it provides the flexibility to use both the file-based approach or the in-memory one by using subclassing.

The second step involved updating the VisIVO Library, a library designed to directly expose the VisIVO Server features within the user code. To take advantage of the new in-memory data handling capabilities, the API of the library was extended.

VisIVO Library is organized using environments, which are represented using a variable that contains all settings for a specific operation. At first, we needed to add a new environment setting to enable the module use the in-memory version of the importer and the viewer classes. 

Furthermore, we implemented a new library function to allow users to retrieve in-memory vectors generated by the importer and pass them to the Viewer. The new function is defined as follows: \texttt{int VV\_SetTableFromImporter(VisIVOViewer* viewer, VisIVOImporter* importer, size\_t tableIndex);}

The first parameter is the \texttt{VisIVOViewer} environment, the second is the \texttt{VisIVOImporter} environment, and the third specifies the index of the desired table to retrieve. This is necessary because each \texttt{VisIVOImporter} can generate multiple tables, for example, if the simulation contains different particle types.

\section{Preliminary Results}

\subsection{VisIVO Results}
The previous preliminary results for VisIVO included the development of a prototype featuring an Importer using Hecuba's API to retrieve the simulated data from a distributed database.

The new results consist of a visualization pipeline built using the VisIVO Library, exploiting the newly implemented features to generate the rendering of the simulation data retrieved using the Hecuba importer without relying on storage.

The pipeline is composed of two steps, one that involves VisIVO Importer and another one involving VisIVO Viewer. 

In the first step, the VisIVO Importer environment is defined and the required attributes are set. Specifically. we define the desired importer, in this case Hecuba, and the "use\_memory" attribute, which instructs the importer to use local memory. These are set using the following two library calls: \texttt{VI\_SetAtt(\&env, VI\_SET\_FFORMAT,"hecuba")} and \texttt{VI\_SetAtt(\&env, VI\_SET\_USE\_MEMORY,"")}. 

Next, the VisIVO Viewer environment is also defined with the required attributes, in this case the particle coordinates and the "use\_memory" attribute. As a final step, we use the new function that allows us to retrieve the data from the Importer, 
\begin{center}\texttt{VV\_SetTableFromImporter(\&viewerEnv, \allowbreak  \&importerEnv, 0)}.\end{center} In this case, it retrieves the first table that has 0 as index, which represents the available gas particles.

We compared the execution time of a pipeline using both the classic file-based approach and the in-memory approach (see Table \ref{tab:visivoperf}). The pipeline consists of two steps: first, the ChaNGa importer processes two Tipsy files containing 2.6 million particles (109 MB) and 161 million particles (6.38 GB); second, VisIVO’s Points Viewer is executed on the gas particles. The file-based approach required 1.34 s and 39.11 s to complete the pipeline on the two datasets, while the in-memory implementation reduced this to 0.815 s and 29.76 s, respectively. In the file-based case, the importer step alone took 0.57 s and 9.86 s. Since the in-memory approach avoids one file write in the importer step and one subsequent read in the viewer step, these performance improvements are consistent with expectations.

\begin{table}[!h]
    \centering
    \begin{tabular}{c|c|c|c}
        \textbf{\# of particles} & \textbf{File-based} &	\textbf{In-memory} &	\textbf{Performance Gain} \\
        \hline
2.60E+06 &	1.34 s &	0.815	& 39\%\\
1.61E+08 &	39.11 s &	29.76	& 24\%\\
    \end{tabular}
    \caption{VisIVO execution time comparison of a pipeline using both the classic file-based approach and the in-memory approach.}
    \label{tab:visivoperf}
\end{table}

We expect to achieve similar results using the VisIVO's Hecuba importer as soon as streaming capabilities are implemented on Hecuba's C++ API.

\subsection{ParaView plugin Results}
\begin{figure}[ht]
    \centering
    \includegraphics[width=\linewidth]{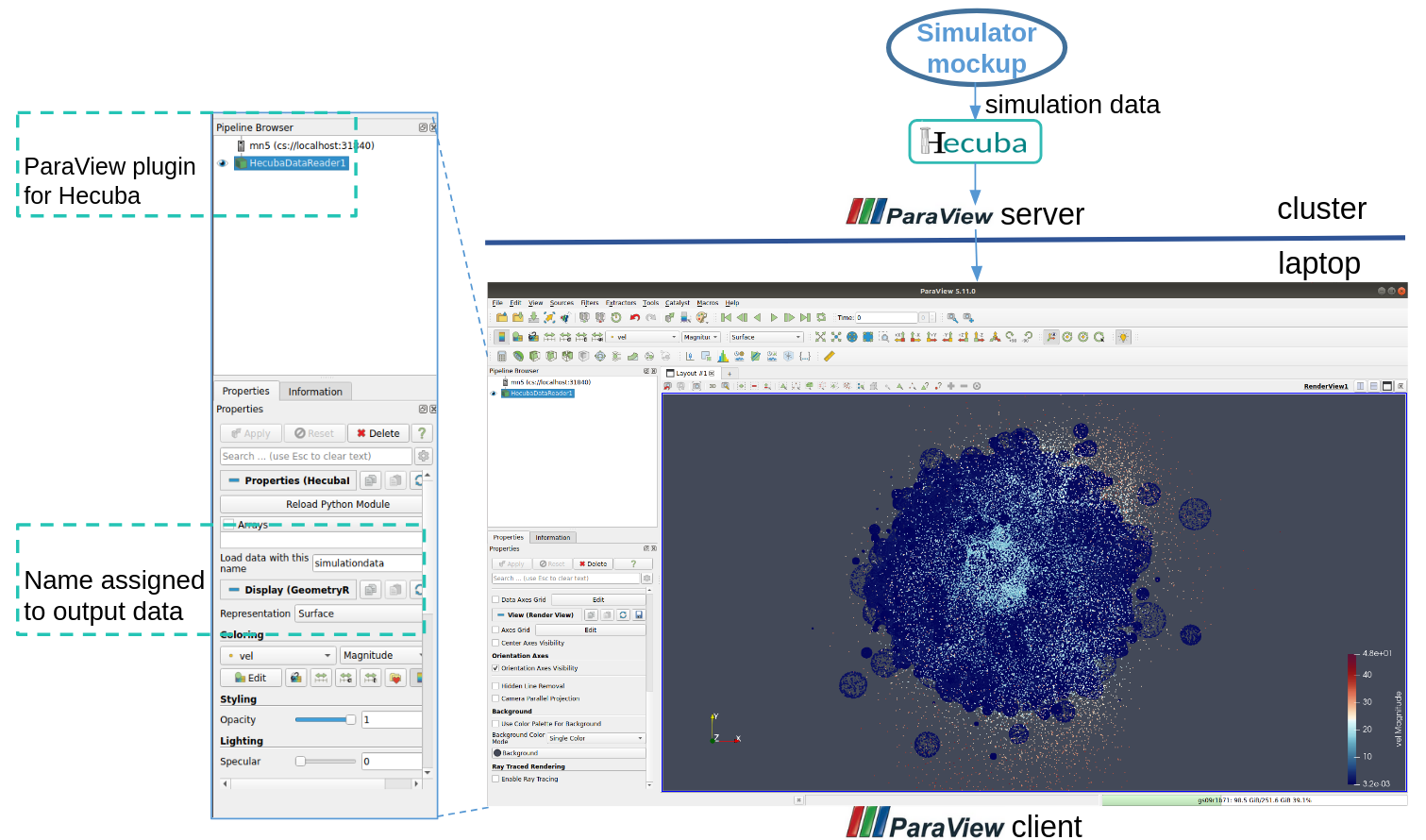}
    \caption{ParaView application window with the plugin with Hecuba integrated that supports in situ analysis.}
    \label{fig:hecuba-plugin}
\end{figure}

Regarding ParaView, we have implemented a Python plugin in which the user is able to request the desired data to read from the stream, browse through all the different timesteps of the simulation and visualize them in the view window as shown in Figure \ref{fig:hecuba-plugin}.
The goal is for the simulator to be able to run on the same infrastructure as the plugin or on a different one. Hecuba's interface allows data to be visualized as it is received without user intervention.
To test the plugin, we implemented a small code that acts as a data generator, as a producer. The generated data is the result of a previous ChaNGa simulation, stored in Tipsy format files. Our producer reads these files and uses Hecuba's interface to send them to our ParaView plugin via stream.

In addition, we have performed preliminary performance evaluations using two Tipsy files, each containing 2.6 million particles (109MB) and 161 million particles (6.38GB).
Our setup offers better performance than if ParaView had to read and manipulate the Tipsy files (see Table \ref{tab:paraviewperf}).
Under these conditions, the times required by the plugin to read from the stream the data from these files using Hecuba are 0.91s and 38.78s, respectively, while the times necessary to read directly from the Tipsy file are 3.02s and 188.3s.

\begin{table}[!h]
    \centering
    \begin{tabular}{c|c|c|c}
        \textbf{\# of particles} & \textbf{File-based} &	\textbf{Hecuba streaming} &	\textbf{Performance Gain} \\
        \hline
2.60E+06 &	3.02 s  & 0.91 s	& 70\%\\
1.61E+08 &	188.3 s &	38.78 s	& 79\%\\
    \end{tabular}
    \caption{Paraview execution time comparison of a pipeline using both the classic file-based approach and the Hecuba streaming approach.}
    \label{tab:paraviewperf}
\end{table}

As part of our next steps, we plan to do a more exhaustive evaluation using more input data sizes. We also aim to implement the interaction of ChaNGa with Hecuba so that it becomes the real producer in our experiments.

\section{Conclusions and future work}

In-situ processing is increasingly adopted to overcome the I/O bottlenecks associated with storing large-scale simulation outputs. This paradigm is especially useful in the context of Astronomy and Cosmology, where the capability to analyze data at runtime enables more efficient workflows.

This paper presented the development and refinement of an in-situ processing strategy leveraging the Hecuba toolset, integrated into a ParaView plugin and the VisIVO visualization environment, to enable run-time in-situ visualization simultaneously with data generation.

For future work in the context of VisIVO, we are working on adding the capability of in-memory data handling also to the VisIVO Filter module. This enhancement will enable more complex workflows, such as the generation of volumetric data using the data received by the Hecuba importer and then performing volume rendering without intermediate storage.

For the ParaView plugin, we are still working to optimize and refactor the code to have a better software pattern and maintenance.
The small code that acts as a producer is a placeholder because we will study how to implement the interaction of ChaNGa with Hecuba to skip the usage of intermediate files.
Also, in addition to the results shown, we aim to perform a more comprehensive evaluation using more input data sizes.

\begin{acknowledgments}
  This work is funded by the European High Performance Computing Joint Undertaking (JU) and Belgium, Czech Republic, France, Germany, Greece, Italy, Norway, and Spain under grant agreement No 101093441 and it is supported by the spoke "FutureHPC \& BigData” of the ICSC – Centro Nazionale di Ricerca in High Performance Computing, Big Data and Quantum Computing – and hosting entity, funded by European Union – NextGenerationEU. 
\end{acknowledgments}

\section*{Declaration on Generative AI}
During the preparation of this work, the author(s) used GPT-4 in order to: Grammar and spelling check. After using these tool, the authors reviewed and edited the content as needed and take full responsibility for the publication’s content.

\bibliography{biblio}

\end{document}